\documentclass[conference]{IEEEtran}
\IEEEoverridecommandlockouts
\usepackage{cite}
\usepackage{amsmath,amssymb,amsfonts}
\usepackage{algorithmic}
\usepackage{graphicx}
\usepackage{textcomp}
\usepackage{xcolor}

\usepackage{hyperref}
\usepackage{listings}
\usepackage{svg}
\usepackage{float}
\usepackage{xspace}
\usepackage{makecell}
\usepackage{caption}

\def\BibTeX{{\rm B\kern-.05em{\sc i\kern-.025em b}\kern-.08em
    T\kern-.1667em\lower.7ex\hbox{E}\kern-.125emX}}
\begin{document}

\newcommand{\RQOne}{What is the precision, recall and F1-score of GPT-5 and GPT-5-mini in code-change impact prediction?}
\newcommand{\RQTwo}{To what extent does the inclusion of the diff hunks of the seed change improve code-change impact prediction?}
\newcommand{\etal}{\textit{et al. }}
\newcommand{\ie}{\textit{i.e., }}
\newcommand{\eg}{\textit{e.g. }}

\newcommand{\figref}[1]{Figure~\ref{#1}\xspace}
\newcommand{\chapref}[1]{Chapter~\ref{#1}\xspace}
\newcommand{\defref}[1]{Definition~\ref{#1}\xspace}
\newcommand{\lstref}[1]{Listing~\ref{#1}\xspace}
\newcommand{\secref}[1]{Section~\ref{#1}\xspace}
\newcommand{\tabref}[1]{Table~\ref{#1}\xspace}

\newcommand{\change}[1]{#1}

%\title{Mind the Change: Code-change Impact Prediction using GPT-5\\
\title{A Dataset and Preliminary Study of Using GPT-5 for Code-change Impact Analysis\\
\thanks{This research was funded in whole or in part by the Austrian Science Fund (FWF) 10.55776/P36698. For open access
purposes, the author has applied a CC BY public copyright license to any author accepted manuscript version arising
from this submission. The research was supported by the Austrian ministries BMIMI, BMWET and the State of Upper
Austria in the frame of the SCCH COMET competence center INTEGRATE (FFG 892418).}
}

 \author{\IEEEauthorblockN{Katharina Stengg}
 \IEEEauthorblockA{\textit{Software Engineering Research Group} \\
 \textit{University of Klagenfurt}\\
 Klagenfurt, Austria \\
 katharina.stengg@aau.at}
 \and
 \IEEEauthorblockN{Christian Macho}
 \IEEEauthorblockA{\textit{Software Engineering Research Group} \\
 \textit{University of Klagenfurt}\\
 Klagenfurt, Austria \\
 christian.macho@aau.at}
 \and
 \IEEEauthorblockN{Martin Pinzger}
 \IEEEauthorblockA{\textit{Software Engineering Research Group} \\
 \textit{University of Klagenfurt}\\
 Klagenfurt, Austria \\
 martin.pinzger@aau.at}
 }
%\author{\IEEEauthorblockN{}
%	\IEEEauthorblockA{\textit{ } \\
%		\textit{ }\\
%		\\
%	}
%	\and
%	\IEEEauthorblockN{Anonymous}
%	\IEEEauthorblockA{\textit{ } \\
%		\textit{ }\\
%		\\
%	}
%	\and
%	
%	\IEEEauthorblockN{}
%	\IEEEauthorblockA{\textit{ } \\
%		\\
%		\\
%	}
%}

\maketitle

\begin{abstract}
Understanding source code changes and their impact on other code entities is a crucial skill in software development. 
However, the analysis of code changes and their impact is often performed manually and therefore is time-consuming. 
Recent advancements in AI, and in particular large language models (LLMs) show promises to help developers 
in various code analysis tasks. 
However, the extent to which this potential can be utilized for understanding code changes and their impact is underexplored. 
To address this gap, we study the capabilities of GPT-5 and GPT-5-mini 
%The time required to understand changes and to
%familiarise oneself with the codebase, for example, during a code
%review, slows down both developers and reviewers. To address
%this challenge, we aim to analyse the capabilities of large language models
%(LLMs) 
to predict the code entities impacted by given source code changes. We
construct a dataset containing information about
seed-changes, change pairs, and change types for each commit.
Existing datasets lack crucial information about seed changes and
impacted code entities. Our experiments evaluate the LLMs in two configurations: 
(1) seed-change information and the parent commit tree and 
(2) seed-change information, the parent commit tree, and the diff hunk of each seed change.
%, and (3) seed-change information and the current commit tree.
%(1) seed-change information alone,
%(2) seed-changes plus minimal Git diff hunks, (3) seed-changes
%plus commit message, and (4) all three together. 
We found that both LLMs perform poorly in the two experiments, whereas GPT-5 outperforms GPT-5-mini. Furthermore, the provision of the diff hunks helps both models to slightly improve their performance.
%shows potential for predicting the impact of code changes with the current commit tree.
%Based on these results, we foresee several directions for future work to improve the use of LLMs for code-change impact analysis.
%Future research can build upon our findings and extend the use of LLMs for code-change impact prediction. 
%Furthermore, our dataset supports future studies, such as identifying co-evolution patterns of Java source code changes.
\end{abstract}


\section{Introduction}
Modern software engineering is based on the frequent adaptation of source code to meet evolving requirements. 
\change{To} make these adaptations, software engineers need to understand the \change{code}, its changes, and the potential
impact of those changes on other entities. %of the codebase
\change{Tao \etal}\cite{HowDoEngineers2012Tao} found that over 43\% of \change{their interviewed software engineers} 
deal with code changes daily, and 36\% even more often.
Furthermore, they found that there is no adequate tool support \change{to assess the quality, understanding, or decomposition of changes.}
\change{A study by Bacchelli and Bird \cite{ExpectationsOutcomesCodeReview2013Bacchelli} focused on modern code review and reported similar findings,
concluding that there is a need for tool improvement and that it is important to understand changes \change{to} 
provide good reviews.}
Recent studies explored the potential of using large language models (LLMs) for automating several code-related \cite{DoAdvancedLLMSEliminate2025Wang}
and code-change related \cite{ExploringLLMsCodeChange2025Fan} tasks,
such as vulnerability \change{detection and repair} \cite{VulRepairLLMs2025Zhou} and
secure code generation \cite{BenchmarkingSecureCodeGenLLMs2025Bruni}.
However, to the best of our knowledge this potential has not been explored for analysing the impact of code changes.
According to Angerer~\etal\cite{ChangeImpactAnalysis2019Angerer}, the change impact denotes the potential effects of a software change.
%generated interest in applying them to improve code-change impact prediction.\\
%Their findings strengthen the need to further investigate the use of LLMs in the SE domain.\\

In this work, we explore the capabilities of two state-of-the-art LLMs, namely GPT-5 and  its smaller version GPT-5-mini, for code change-impact prediction. 
Furthermore, we extend the dataset \textit{ALEXANDRIA} from Yan \etal \cite{EnhancingCodeUnderstanding2024Yan},
as their dataset contains only changed method pairs and does not consider 
attribute or class changes.
%extend an existing dataset to provide not only information on the code changes but also classifies their type 
%and relationship to other changes within the same commit.
%present a novel benchmark, that 
%We aim to tackle automated code-change impact understanding by using state-of-the-art LLMs for code change-impact prediction. Furthermore, we propose a novel benchmark, that 
%provides not only information on the code changes but also classifies their type 
%and relationship to other changes within the same commit.
%also lists the seed-changes and change pairs to serve
%as an accurate ground truth set in this domain. \\
%This study highlights the need for novel approaches for change-impact prediction
%as the delay due to time need to understand changes impacts not only the reviewer but also the developer. \\
%We perform four distinct experiments to use our dataset and evaluate our idea and answer 2 research questions.
For our preliminary evaluation, we performed two experiments. The first experiment, named \textit{Basic}, provides the LLM with the link to the parent commit tree and the fully qualified names of the seed changes. 
A seed change denotes the code changes that trigger changes in other code entities. 
The second experiment, named \textit{Diff}, extends the first experiment by providing the diff hunks of the seed changes. 
%
%All experiments use information about either the respective parent commit tree or the current commit tree, together with the fully 
%qualified name of the seed changes in the current commit.
%The goal of each experiment is to compare the capabilities of GPT-5 and 
%GPT-5-mini, given this information about seed changes, to predict the set of impacted code entities.\\
%We study different prompts per experiment.
%The first experiment, \textit{Basic}, is compared with the second experiment, \textit{Diff}, and the third experiment, \textit{Curr},
With the two experiments, we aim to answer the following two resesearch questions:
\begin{itemize}
\item RQ1: \RQOne
\item RQ2: \RQTwo
\end{itemize}

%to answer \textbf{(RQ1)} \textbf{\RQOne}
%
%The experiment \textit{Basic} provides the LLMs with a link to the parent commit tree, the experiment \textit{Diff} extends Basic by providing the LLMs with the diff hunk of the seed changes.
%The third experiment, \textit{Curr}, provides the the commit tree of the commit that contains all changes.
%The experiment \textit{Diff} uses a prompt with additional information in form of the minimal Git diff hunk of the seed-changes. \\
%To investigate whether LLMs benefit from the knowledge of commit messages, 
%we extend experiment \textit{Basic} with this information, the corresponding experiment is called \textit{CM}.
%The fourth experiment, \textit{Diff \& CM}, combines all previous experiments, using the fully qualified names of the seed changes,
%the Git diff hunk, and the commit message for code-change impact prediction.
%These two experiments are used to answer \textbf{(RQ2)} \textbf{\RQTwo}\\
The results with 40 commits randomly selected from popular Java projects on GitHub show that 
GPT-5 outperforms GPT-5-mini in both experiments. %, except \textit{CM}.
LLMs benefit from knowing the commit that contains the changes. % intent and description
In addition, the provision of the diff hunks helps both models to slightly improve their performance.
%but performance decreases when further information, \eg the minimal diff hunk is provided.}
%This can be due to confusion of the models.\\
%
This paper makes the following contributions:
\begin{enumerate}
	\item The \textit{Alextend} dataset, which contains detailed information about code changes and their impact.
	\item First results of the performance of GPT-5 and GPT-5-mini for code-change impact prediction.
\end{enumerate}

\section{Related Work}
%\subsection{Background}
Hanam \etal \cite{AidingCodeUnderstanding2019Hanam} proposed a tool called \textit{SEMCIA}, 
which improves the code reviewing process for developers in terms of time and performance.
Furthermore, they defined semantic change relations and performed a user study comparing traditional change impact analysis techniques with their proposed approach.
In 2023, Lin \etal \cite{CCT52023Lin} proposed a model based on the Text-To-Text-Transfer Transformer (T5) model. 
Their model is pre-trained on five tasks. For evaluation, they fine-tuned it for each of the five code-change related tasks. 
These tasks are commit message generation, just-in-time comment update, and defect prediction as well as 
code-change quality and code review generation.
They proposed and used a large-scale dataset called CodeChangeNet.
This dataset consists of 
code-change and commit message pairs of six programming languages. 
In their work \cite{EnhancingCodeUnderstanding2024Yan}, Yan \etal
propose their impact analysis approach \textit{ATHENA}, which is based on GraphCodeBERT and structural dependence graphs.
They also provide a novel code-change impact analysis dataset
called \textit{ALEXANDRIA}.
This dataset contains method-level Java production code changes and meta-information.
Compared to our dataset, \textit{ALEXANDRIA} indicates no clear seed method, therefore they perform 
their impact analysis experiments treating all method-level changes of a commit as query and impact.
The paper also states that this may not be a realistic scenario. This is a gap that we address in this work.

In their work \cite{ExploringLLMsCodeChange2025Fan}, Fan \etal compare various 
LLMs and small-pretrained models (\eg, CCT5 \cite{CCT52023Lin}) for 
code review generation, commit message generation and just-in-time comment update. 
They investigated whether in-context learning (ICL) or parameter-efficient fine-tuning (PEFT) improves performance,
and compared two input formats, the code itself and the code diff, aiming to find the best combinations per code-change task.
%as well as only document changes, code changes or a combination. 
The authors state that LLMs offer potential for code-change related tasks, but did not evaluate this potential for code-change impact analysis.

%\section{Approach}
%This section describes (1) datset generation, (2) LLM selection, (3) prompt definition, and (4) change impact prediction.
\section{Fine-Grained Change Impact Dataset}
Our dataset \textit{Alextend} is an extension of the \textit{ALEXANDRIA} dataset proposed by Yan \etal \cite{EnhancingCodeUnderstanding2024Yan}.
The original dataset contains method-level changes and corresponding information, such as 
the repository name, commit hash, parent commit hash, file path, etc. per changed method.\\
From \textit{ALEXANDRIA}, we randomly selected 40 commits from all Java projects that satisfied the following criteria: 
\begin{itemize}
	\item A minimum of one and a maximum of five changed Java source code files 
	\item A minimum of two changes, that are not comment, import, or systematic changes (identical modifications across multiple code entities)
\end{itemize}

\begin{table}[htbp]
\caption{Descriptive statistics of the 40 commits.}
\begin{center}
\begin{tabular}{lrr}\hline
\textbf{Project} & \textbf{\#Commit} & \textbf{\#Change pair} \\\hline
apache/ant-ivy & 12 & 66 \\
apache/commons-codec & 1 & 4 \\
apache/commons-compress & 3 & 4 \\
apache/commons-configuration & 4 & 23 \\
apache/commons-io & 1 & 2 \\
apache/commons-lang & 1 & 2 \\
apache/commons-math & 3 & 12 \\
apache/commons-net & 3 & 15 \\
apache/commons-scxml & 1 & 1 \\
apache/commons-vfs & 4 & 10 \\
apache/giraph & 2 & 6 \\
apache/gora & 1 & 8 \\
apache/opennlp & 3 & 6 \\
apache/parquet-mr & 1 & 33 \\\hline
Total & 40 & 192 \\\hline
\end{tabular}
\label{tab:project_statistics}
\end{center}
\end{table}

\tabref{tab:project_statistics} lists the number of selected commits (\textit{\#Commit}) and the number of manually identified change pairs (\textit{\#Change pair}) per project.
In total, our dataset contains 40 commits with 192 code-change pairs that spread among 14 projects.
%\tabref{tab:project_statistics} lists our dataset's projects with their descriptive statistics.
%\textit{\#Commit} refers to the amount of commits per project and \textit{\#Change pairs} counts the change pairs per project.

For each commit, our dataset stores the attributes \textit{repo}, \textit{commit\_hash}, and \textit{parent\_commit\_hash},
that list the repository name and the hash of the commit and its parent commit.
The \textit{github\_link} attribute provides the URL of the repository at the current commit on GitHub.
Furthermore, the dataset contains the field \textit{java\_class\_count}, which represents the amount of changed Java source code files 
per commit. 

Compared to \textit{ALEXANDRIA}, we include not only method, but also class and attribute changes. We first extracted all Java source code changes that do not alter comments or imports per commit.
We use the fully qualified names\footnote{\url{https://docs.oracle.com/javase/specs/jls/se17/html/jls-6.html\#jls-6.7}} to make our code entities uniquely identifiable.
For example, the changed method \textit{foo(int)} is identified as 
\textit{org.test.Class1.foo(int)}. We do not use fully qualified names for method parameters 
for improved readability and reduced token count. %, and because the dataset does not contain cases where methods would not be uniquely identified. 
If a nested class declaration or body was changed, the fully qualified name includes
both the enclosing and the nested class, \eg \textit{org.test.Class1\$nestedClass.method1()}.
%Two co-authors independently annotated the code changes of each selected commit.

Furthermore and compared to \textit{ALEXANDRIA}, for each commit, our dataset contains 
the type of each change, the change pairs, the seed changes, and the diff hunk for the seed changes. 
Regarding the change types, change pairs, and seed changes, two co-authors independently
and manually annotated the code changes of each commit.
Thereafter, they met and discussed the results until they reached agreement on all annotations. 
In the initial phase, seed changes were often misinterpreted, therefore we defined them as the primary logical changes that trigger the remaining changes.

%\textbf{Change Types. }
%\subsection{Change categories}
For the annotation of the \textit{change types}, the two co-authors used the change type taxonomy 
introduced by Fluri and Gall~\cite{ClassifyingChangeTypes2006Fluri}.
Their taxonomy distinguishes between declaration and body changes. 
The main categories for classes are \textit{Class Declaration Changes} and \textit{Class Body Changes}.
Class body changes can be split up into \textit{Method Declaration Changes}, \textit{Method Body Changes}, and \textit{Attribute Declaration Changes}.
We skipped the types \textit{Access Modifier Changes} and \textit{Final Modifier Changes}. 
%We used these categories except the last two, aiming for a coarse-grained classification to get an overview of the change types.
A method body change is also a class body change, but we recorded it solely as method body change as we 
classified changes only by their most specific type.
The results are stored in the column \textit{change\_categories} of our dataset.

%\tabref{tab:change_categories} lists the change categories and their occurence counts.
%In total, we found 258 changes in the 40 commits.
%117 method body changes, 89 method declaration changes, 44 attribute declaration changes,
%and 7 class declaration changes. All class declaration changes are due to new classes.
%One class body change was found.
%This is the change of an enum class value, which results in a change of the enumeration definition, \ie no method nor attribute changed.
%Furthermore, the authors provide a fine-grained taxonomy, where they state change categories for each of these. 
%We stick with the coarse-grained categorisation, as change categories are not the main objective in this work. 
%\begin{table}[t]
%	\caption{code-change categories and their count}
%	\begin{center}
%		\begin{tabular}{lr}\hline
%			\textbf{Change Type} & \textbf{\#Count} \\
%			\hline
%			MethodBodyChange & 117 \\
%			MethodDeclarationChange & 89 \\
%			AttributeDeclarationChange & 44 \\
%			ClassDeclarationChange & 7 \\
%			ClassBodyChange & 1 \\\hline
%			Total & 258\\\hline
%		\end{tabular}
%		\label{tab:change_categories}
%	\end{center}
%\end{table}

%\textbf{Seed-changes and Change Pairs. }
%\subsection{Change pairs}
%\subsection{Seed-changes}
For the annotation of the \textit{seed changes} and \textit{change pairs},
the two co-authors analysed each code change for its relationship to the other changes. 
\lstref{lst:motivating_example} shows an example of a change in Java code.
The class attribute \textit{loadCounter} has been changed from \textit{int} to \textit{AtomicInteger}.
Because an AtomicInteger must be initialised before it can be used, the constructor creates a new instance.
\textit{loadCounter} and \textit{PropertiesConfigurationLayout(PropertiesConfigurationLayout)} form a change pair, whereas 
\textit{loadCounter} is a seed-change, \ie the change that triggers the change in the constructor. 
This dependency is of semantic nature rather than of syntactic nature, making it hard for existing approaches to capture.
Our dataset stores the change pairs of each commit as a semicolon-separated list in \textit{change\_pairs}
and the seed changes in \textit{seed\_changes}.

%Our dataset records both syntactic and semantic change relationships, so we annotate them manually rather than using automated tools.
%\begin{figure}[t] % * to span over both columns
\begin{lstlisting}[language=Java,caption={Code-change impact example},
	label={lst:motivating_example},
	linewidth=\columnwidth]
@@ -133,7 +134,7 @@ public class PropertiesConfigurationLayout
...
-	private int loadCounter;
+	private final AtomicInteger loadCounter;
...
public PropertiesConfigurationLayout(PropertiesConfigurationLayout c)
{
+		loadCounter = new AtomicInteger();
\end{lstlisting}
%\end{figure}

%\textbf{Git Diff Hunk. }
Our dataset also contains the minimal \textit{diff hunk} of each seed change of a commit.
The minimal diff hunk is exported from GitHub and consists of:
\begin{itemize}
	\item An identifier of the changed entitiy;
	\item Lines starting with + that show added code;
	\item Lines starting with - that show removed code;
\end{itemize}

We removed any further diff information to save input tokens when using the diff hunk in our prompts.
The git diff hunk is stored in the column \textit{seed\_changes\_diff}.
Finally, we also stored the commit message of each commit in the column \textit{commit\_message}.
%our dataset contains argumentation on the labelling (\textit{argumentation}) and the corresponding commit message (\textit{commit\_message}).\\
%The following parts make up our dataset:
%\textit{sample} is the respective number of the sample in the original, shuffled, and grouped dataset. 
%\textit{repo} consists of two parts seperated by \, namely the username of the owner followed by the repository name of the corresponding GitHub repository. 
%\textit{commit} is the commit hash of the current commit, \textit{parent\_commit} is the commit hash of the commit's parent. 
%\textit{java\_class\_count} is the amount of Java source code files that were changed at the current commit. 
%\textit{change\_categories} lists each source code-change and its corresponding change category. 
%\textit{change\_pairs} is a semicolon-seperated (;) list of pairwise-connected changes. 
%For example, if a method siganture changed and the corresponding usage also changed, these two changes form a change pair. 
%\textit{seed\_changes} is a semicolon-seperated list of the seed-changes that triggered the remainder of the 
%changes. 
%\textit{argumentation} provides a natural language description of the choices during the labelling process.
%\textit{seed\_changes\_diff} is the minimal diff hunk of the seed-changes. \\
%\\

\section{Prompt Definition}
\lstref{lst:prompt_optimized} shows the basic prompt created for our experiments. 
We followed guidelines and best practices
from other research papers and OpenAI \cite{OpenAI2025Guidelines}.
In their work \cite{InContextImpersonation2023Salewski}, Salewski \etal found that telling an LLM to act as a domain-expert can lead to better results. 
Therefore, all our prompts start with the impersonation of the LLM as an intelligent assistant that helps with the analysis of Java source code changes and their impact on other code entities. 
The prompt then states instructions and rules, as LLMs benefit from precise information. 
The specified input variables \textit{github\_link\_parent} and \textit{seed\_changes} are dynamically updated during the experiments. 
Finally, we define the JSON output format and provide two examples, one with an impacted method and attribute and
one with no impacted entities.
%\begin{figure}[t]
	%\caption{Basic prompt optimized for GPT-5 models}

\begin{figure}[]
\begin{lstlisting}[caption={Basic prompt optimized for GPT-5 models}, 
	label={lst:prompt_optimized},
	linewidth=\columnwidth]
You are an intelligent assistant designed to help developers and code reviewers analyze Java source code changes and assess their impact on related code entities.

## Instructions
You will be provided with:
- A link to a parent commit in a GitHub repository.
- A list of seed changes made to the code.
Your task is to:
- Identify every Java code entity affected by these seed changes; specifically, those entities that must be modified as a direct result of the seed changes.

Follow these rules strictly:
1. For each impacted code entity, list its fully qualified name (FQN). For nested or inner classes, use the fully qualified name of the outer class, followed by a dollar sign (`$`), then the inner class name, and then the changed entity (e.g., `com.example.Outer$Inner.changedEntity`).
2. Do not use fully qualified names for method parameters; only for Java classes, interfaces, methods, attributes, and variables.
3. Do not include the seed changes themselves in the output; any entity listed as a seed change must be excluded.  
4. Exclude any changes to test files, `.md` files, or `.xml` files; your analysis should focus exclusively on Java source code files.
5. You must only view and analyze the parent commit provided; reviewing or utilizing information from any follow-up or subsequent commits is strictly forbidden.
6. If a code entity is renamed due to the seed changes, list only the new fully qualified name in your output.
7. Impacted entities must be those that must change *because of* the seed change, not the seed change itself.

If ambiguity or uncertainty occurs, do not ask for clarification; choose and report the entities most likely to be impacted.

## Inputs
- **Commit link:** {{github_link_parent}}
- **Seed changes:** {{seed_changes}}

## Output Format
Return a JSON object with a single property named "impacted_entities", whose value is an array of strings. Each string should be the fully qualified name of an impacted code entity. If there are no impacted code entities, return an empty array for "impacted_entities".

**Example:**
```json
{"impacted_entities": ["com.example.Foo.someMethod(String, int)", "com.example.bar.attribute1"]}
```

**Example with no impacted entities:**
```json
{"impacted_entities": []}
```
\end{lstlisting}
	%\label{lst:prompt_optimized}
%\end{figure}
\end{figure}

Wang \etal \cite{DoAdvancedLLMSEliminate2025Wang} tested prompt engineering techniques for code-related tasks on various models. 
They provided guidelines for the choice of LLM and prompt engineering technique
and highlighted the importance of adapting the prompt to the strengths and weaknesses of the chosen LLMs. 
To optimize our prompt for the use with GPT-5 and GPT-5-mini, the prompt optimizer from OpenAI was used\footnote{\url{https://platform.openai.com/chat/edit?models=gpt-5&optimize=true}}.
%\section{LLM Selection}
%\url{https://openai.com/index/introducing-gpt-5-for-developers/?utm_source=chatgpt.com} da sagens 5 is strongest fia coding 
%\url{https://platform.openai.com/docs/guides/latest-model#gpt-5-parameter-compatibility}

\section{Preliminary Evaluation}
This section describes the preliminary evaluation of our proposed approach.
We used our dataset \textit{Alextend} and the models GPT-5 and GPT-5-mini
to evaluate their performance in predicting the impact of given code changes. 
%
%We chose the LLM \textit{GPT-5} because its provider \textit{OpenAI} states, that GPT-5 is their best model for coding\footnote{\url{https://platform.openai.com/docs/models/gpt-5}}.
%Our experiments compare GPT-5 with its smaller and cheaper version \textit{GPT-5-mini}\footnote{\url{https://platform.openai.com/docs/models/gpt-5-mini}}.
The precise versions of the models are \textbf{gpt-5-2025-08-07} (GPT-5) and \textbf{gpt-5-mini-2025-08-07} (GPT-5-mini). 
GPT-5 was trained on information up until Sep 30, 2024 and GPT-5-mini up until May 31, 2024.
Both models have a context window of 400,000 tokens. 
As it is practically not feasible to equip our prompts with the whole codebase of a Java project, we decided to enable the tool web search\footnote{\url{https://platform.openai.com/docs/guides/tools-web-search?api-mode=responses}},
such that the models can look up the provided commit tree on GitHub.

The models' responses are compared with our labelled ground truth impact set computed from the change pairs in our dataset. 
More specifically, the ground truth impact set contains all direct and indirect dependent code entities of the seed changes in a commit.
%These changes are transitively connected to the seed change through the change pairs.
For each response, we calculate precision, recall, and F1-score per commit based on exact matches of the fully qualified names of impacted code entities. 
We implemented our experiments in a Python script using the \texttt{openai}\footnote{\url{https://github.com/openai/openai-python}} library. 

\subsection{Using the Basic Prompt}
For answering RQ1, our Python script iterated over the 40 commits in our dataset and for each commit composed and sent the \textit{Basic} prompt 
shown in \lstref{lst:prompt_optimized} to GPT-5 and to GPT-5-mini. 
This prompt contains the link to the parent commit tree and the fully qualified names of the seed changes.
The responses were stored in a Hugginface dataset. In addition, logs of 
responses were stored to files for subsequent analysis. 

%\subsubsection{Setup}
%The first experiment (\textit{Basic}) uses the prompt shown in \lstref{lst:prompt_optimized} and provides the LLMs with a link to the parent commit tree on GitHub
%and the fully qualified names of the seed-changes. 
%The third experiment (\textit{Curr}) uses the prompt of the basic experiment but provides the LLMs with a link to the commit tree of the commit containing the changes.

\begin{figure*}[htb]
	\centerline{\includegraphics[width=0.8\linewidth]{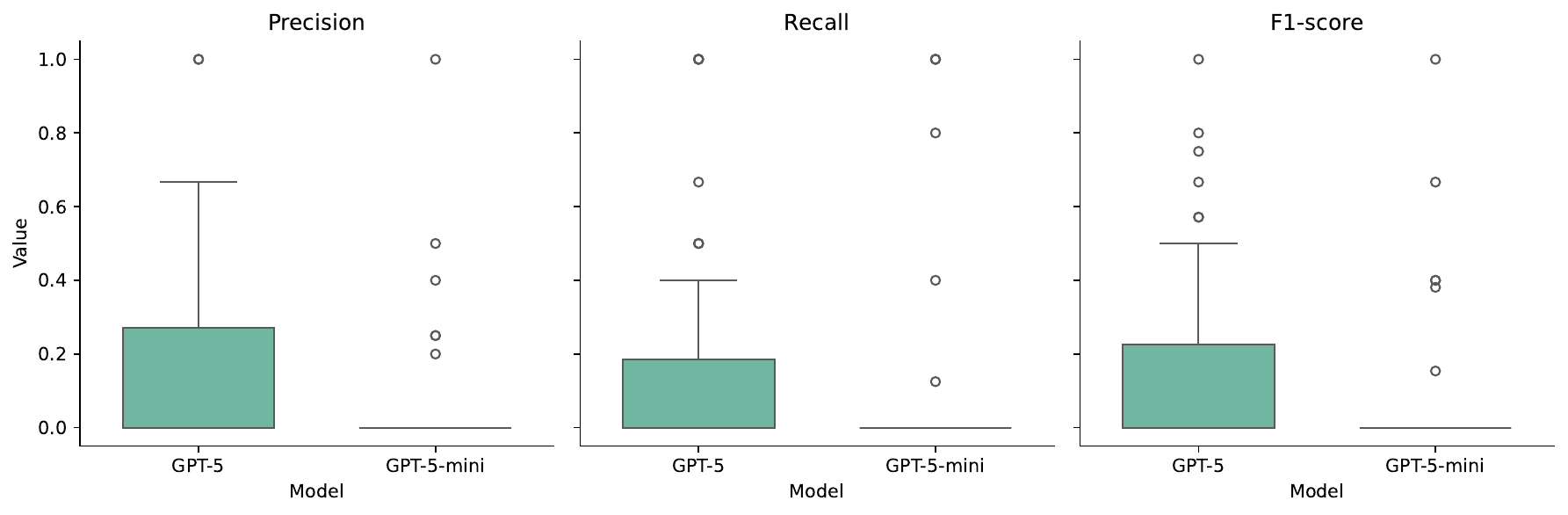}}
	\caption{Precision, recall, and F1-score of all 40 commits per model for the experiment \textit{Basic}}
	\label{fig:boxplot_rq1}
\end{figure*}

%\subsubsection{Results}
\figref{fig:boxplot_rq1} visualises the precision, recall, and F1-score of experiment \textit{Basic} for the 40 commits of the dataset.
%Experiment one, \textit{Basic}, outperforms experiment two
%\tabref{tab:results_rq1} shows the micro- and marco-average values of precision, recall, and F1-score per model and experiment.
%The highest values per metric are marked in bold and highlight, that 
%GPT-5 outperforms GPT-5-mini in both experiments. 
%This performance differences can be due to the lack of smaller models to handle more complex prompts and tasks.

Overall, the box plots show that both models perform poorly in this setting, with median values of 0.0 for precision, recall, and F1-score.
Comparing the plots for GPT-5 and GPT-5-mini, we see that GPT-5 outperforms GPT-5-mini in all three metrics. 
The third-quartile scores (25\% of the tested samples) of GPT-5 show 0.27 for precision, 0.18 for recall, and 0.22 for F1-score compared to 0.0 for all three metrics for GPT-5-mini 
%mean that 25\% of the tested samples achieved values above these thresholds.
%GPT-5-mini recorded a third-quartile of 0.00 for all metrics.
%We consider a prediction as a true positive only if it matches the fully qualified name of the changed entitiy. 

\subsection{Adding Diff Hunks}
The second experiment aims to answer RQ2 and extends the \textit{Basic} prompt with the minimal diff hunk of the seed changes in a commit, as shown in \lstref{lst:prompt_diff}.

\begin{figure}[H]
\begin{lstlisting}[caption={Adding the diff hunks of seed changes to the prompt for experiment \textit{Diff}.}, %
	label={lst:prompt_diff},
	linewidth=\columnwidth]
	- The code diff from git (just the minimal hunk: identifier of changed entity and changed lines (add is indicated via +; delete via -)).
	...
	- **Code diff:** {{seed_changes_diff}}
\end{lstlisting}
\end{figure}

\begin{figure}[]
	\centerline{\includegraphics[width=0.8\linewidth]{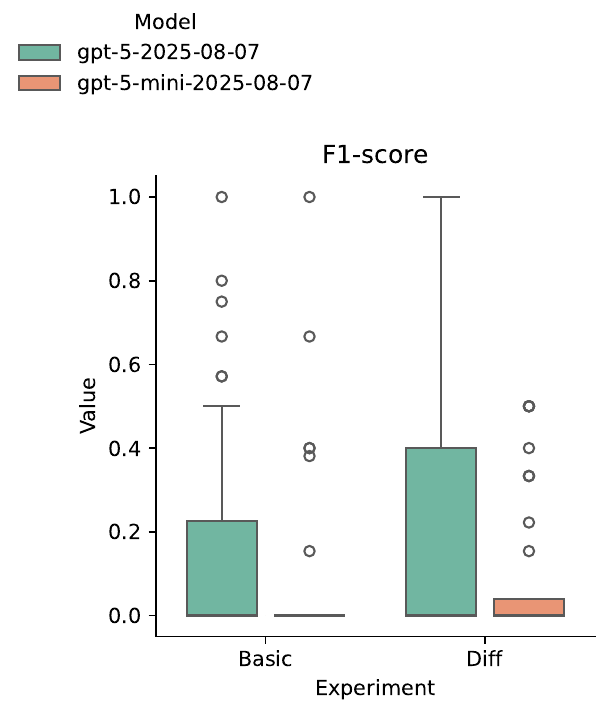}}
	\caption{F1-score of all 40 commits per model for the experiments \textit{Basic} and \textit{Diff}}
	\label{fig:boxplot_rq2}
\end{figure}

\figref{fig:boxplot_rq2} shows the box plots of the F1-scores of GPT-5 and GPT5-mini and compares them with the F1-scores from experiment \textit{Basic}.
For experiment \textit{Diff}, both models again achieved a median F1-score of 0.00. But compared to the \textit{Basic}, the values for the third-quantile scores
increased for both models. GPT-5 achieved a Q3 F1-score of 0.4, while GPT-5-mini achieved 0.04.
In comparison, in experiment \textit{Basic}, GPT-5 achieved a F1-score of 0.22 and GPT-5-mini of 0.00.
These results indicate that code-change impact prediction improves when the LLM is not only provided with the fully qualified name of the changed code entity
but also with the information about the code changes in the form of diff hunks. 

%GPT-5 outperforms GPT-5-mini in both experiments. 
%T%he performance differences of the two models can be due to smaller models having difficulties in handling more complex prompts and tasks.
%Furthermore, we consider a prediction as a true positive only if it matches the fully qualified name of the changed entitiy. 
%For instance, if the model predicts the correct impacted function but includes a wrong parameter or fully qualified name, it is not considered as correct prediction. 

%\begin{figure}[t]
%	\centerline{\includegraphics[width=\columnwidth]{Research_Prototype/code/final_results/boxplot_rq2.pdf}}
%	\caption{F1-score of all 40 commits per model for the experiments \textit{Basic} and \textit{Diff}}
%	\label{fig:boxplot_rq2}
%\end{figure}

\subsection{Log Analysis}
For each experiment, we analysed the logs, focusing on the web search calls of the two models to gain further insight into web information retrieval.
Overall, the logs show that for each prompt, both models performed various web search tool calls to different websites.
We also found that neither model queried the current commit or the code changes in it. 
They only accessed the source code denoted by the parent tree, \ie they followed the instructions provided in the prompt. 

For experiment \textit{Basic}, GPT-5 always queried at least once a folder at the parent tree and GPT-5-mini did so for 75\% of the tested commits.
GPT-5 wrongly queried the parent commit, which contains the code changes of the previous commit, in two samples. 
For experiment \textit{Diff}, GPT-5 always queried at least once a folder at the parent tree except for one sample, where it neither queried the parent tree nor any folder.
GPT-5-mini queried no parent tree for four commits, and did not query any folder of the parent tree for 22 of the 40 tested commits.
This partially explains the poorer performance of GPT-5-mini.

%In comparison to GPT-5-mini, GPT-5 frequently queries the current commit, \ie the url \url{https://github.com/owner/repository/commit/commit_hash}.
%This behaviour may be caused by the prompt, which stated that a link to the commit in a GitHub repository would be supplied, but did not specify that the link actually points to the commit's tree.
%The websearch calls of the experiment \textit{Curr} indicate, that the improved performance metrics of experiment \textit{Curr} were achieved by looking at the actual commit rather than understanding the project's code.

\subsection{Costs}
The costs per 1M tokens for GPT-5 are \$1.25 for input, \$0.125 for cached input, and \$10.00 for output. 
In comparison, 
1M input tokens using GPT-5-mini cost \$0.25, cached input tokens \$0.025, and output tokens \$2.00.
1k web search tool calls costs \$10.
\tabref{tab:costs} lists the costs of the two experiments presented in this paper in US dollars (\$).
%These values are exported from the \textit{Usage} tab on the OpenAI page\footnote{\url{platform.openai.com}}.

We observe that the web search tool is the main cost factor, with a total cost of \$20.91 for 2,091 calls. 
%Another observation is that with \$19.54, the use of the model GPT-5
%costs almost double the price of GPT-5-mini (\$10.21). %When web search is enabled, this cost factor shrinks to about 2.45.
Another observation is that using GPT-5 costs almost twice the price of GPT-5-mini.
The two experiments resulted in a total cost of \$29.75.
\begin{table}[H]
	\caption{Total costs of the experiments per model in US dollars (\$)}
	\begin{center}
		\begin{tabular}{lccccc}
		\hline
		\multicolumn{1}{l|}{\textbf{Model}}&\multicolumn{4}{c}{\textbf{Price (\$)}}\\
		%\cline{1-6} 
		\multicolumn{1}{l|}{\textbf{}}& Input & Cached Input & Output &Web Search & Total\\\hline
		%& & \multicolumn{3}{c}{\textbf{Micro-average}}\\\hline
		%GPT-5-mini &   & 0.37 & 0.26 & 0.30 \\
		\multicolumn{1}{l|}{GPT-5} 			&4.41	& 0.04 	& 3.30 	& 11.79	& 19.54\\
		\multicolumn{1}{l|}{GPT-5-mini}  	& 0.68	& 0.01  & 0.40  & 9.12& 10.21\\\hline\hline
		%\multicolumn{1}{l|}{Web search}  	& - 	& - 	& - 	& 25.97\\\hline 
		\multicolumn{1}{l|}{Total}     		& 5.09 	& 0.05	& 3.70	& 20.91 & 29.75\\\hline
		\label{tab:costs}
		\end{tabular}
	\end{center}
\end{table}

\subsection{Threats to Validity}
There is a potential bias in the experiments due to the uncertainty of the training data of the LLMs. 
We can not be sure whether they already used the selected repositories in their training. 
Similarly, our sample of 40 commits might not be representative. 
We plan to address both threats in our future work.
%We mitigitated this risk by testing our approach on commits made after the model's last training date.
Another threat of this approach is the use of the OpenAI web search feature. 
We mitigated this threat by analyzing the logs with a focus on the links queried by web search. 
%We found only a few cases in which web search queried the commit information that might poin
%Without extracting and validating the logs, we cannot be sure which pages the LLM viewed during the experiments.
%To elict this threat, we plan to evaluate the logs of our experiments in more detail and to test local, sandboxed coding agents.\\
%
Non-determinism of LLMs is another potential threat to the validity of our results. 
To mitigate the risk, we plan on repeating our experiments ten times for evaluation.
The dataset was labelled by two co-authors, which could introduce bias. 
We mitigated this bias by discussing all differences in the labeling until consensus was reached. 
%
%We plan on calculating the margin of error using an unbiased evaluator. 
%This confidence intervall quantifies the labelling errors. 
We used commits with a maximum of five changed Java source code files. 
This can affect our dataset, as we leave out certain commits. But on the other hand, this reduced the bias introduced by very large changes that span multiple change requests.
Another potential threat concerns the prompt itself, as slightly different instructions, for example the change of one word in the prompt, can lead to unforeseen results. 

%For instance, we indicated that the model will receive a link to a commit on GitHub, but did not specify that the link actually 
%pointed to the commit's tree. We plan to test various prompt formats to elict this threat.
%Our ground truth impact set are all changes, that are part of our set of change pairs.
%We therefore assume, that each pair is transitively connected to the seed-change.
%This must not hold for all examples and therefore is a potential threat. 
%We plan on refining our labelling process to mitigate this risk.
%The experiments including the commit message are potentially biased due to the commit message
%being written after seed-changes and impacted changes were made. 
%To reduce this bias, we plan on performing detailed analysis of the commit messages in the dataset.
%Furthermore, we plan on an additional experiment, where the LLM only knows the parent commit's codebase and the commit message.

\section{Conclusions and Future Work}
We presented a preliminary study of the capabilities of GPT-5 and GPT-5-mini for code-change impact prediction. 
For that, we also curated the novel dataset \textit{Alextend}, that is built upon 40 commits 
from the \textit{ALEXANDRIA} dataset \cite{EnhancingCodeUnderstanding2024Yan}. 
Compared to \textit{ALEXANDRIA}, our dataset provides code changes, their type, change pairs, and information about the commit's 
seed-change as well as the corresponding commit message. % and an argumentation about the labelling decisions.
Using our dataset, we experimented with two prompts, whereas the \textit{Basic} prompt contains the fully qualified name of the seed change and the \textit{Diff} prompt adds the
diff hunks. In both settings, the results show that GPT-5 and GPT-5-mini performed poorly, whereas GPT-5 outperformed GPT-5-mini.
Furthermore, the provision of the diff hunks showed to help both models to slightly improve their performance. 
%Our experiments using GPT-5 and GPT-5-mini in two different settings provide an initial direction for using the dataset.

%We considered only exact matches of the fully qualified names of changed code entities as correct.
In ongoing and future work, we first plan to perform our experiments with recent coding agents, such as Claude Code\footnote{\url{https://www.claude.com/product/claude-code}} or Codex\footnote{\url{https://openai.com/codex/}}. 
Furthermore, we plan to integrate sophisticated code analysis tools to provide the models with control and data flow information.
Finally, we plan to extend our dataset to contain the information from more commits. 

%Furthermore, we envision to analyse the relationship of the change types to 
%either link them to taxonomies for code-change co-evolution or to propose an improved taxonomy.
%We aim to evaluate LLMs and state-of-the-art coding agents using different types of prompts and input, such as data and control dependencies,
%to improve the landscape of AI for code-change impact prediction.

%In future work, we plan to evaluate partial matches using alternative strategies
%to evaluate partial matches to improve our understanding of the predictions and results.
%%Our experiments using GPT-5 and GPT-5-mini in two different settings show the potential of the use of LLMs for code-change impact prediction. 
%Furthermore, we plan on improving the dataset, in both terms of quality and size, and to test our proposed approach on a larger dataset consisting of 400 samples instead of 40. 
%Furthermore, we envision to analyse the relationship of the change types to 
%either link them to taxonomies for code-change co-evolution or to propose an improved taxonomy.
%We aim to evaluate LLMs and state-of-the-art coding agents using different types of prompts and input, such as data and control dependencies,
%to improve the landscape of AI for code-change impact prediction.

%As research \cite{DoAdvancedLLMSEliminate2025Wang} highlights the importance of
%input information, 
%we aim to investigate the impact of different inputs, such as data \& control dependencies on code-change impact prediction.

\end{document}